\documentclass[twocolumn,prl,aps,superscriptaddress]{revtex4}
\usepackage{amsmath}
\usepackage{graphicx}
\usepackage[usenames]{color}

\usepackage[ps2pdf,colorlinks,bookmarks]{hyperref}
\definecolor{linkblue}{rgb}{0,0,0.8}
\definecolor{linkgreen}{rgb}{0,0.5,0}
\hypersetup{pdfpagemode=None, pdfstartview=FitH, linkcolor=linkblue, %
             citecolor=linkgreen, urlcolor=linkblue}

\def\beq{\begin{equation}}
\def\eeq{\end{equation}}
\def\bea{\setlength\arraycolsep{1.4pt}\begin{eqnarray}}
\def\eea{\end{eqnarray}}
\def\bit{\begin{itemize}}
\def\eit{\end{itemize}}

\def\eq{Eq.~}
\def\eqs{Eqs.~}
\def\fig{Fig.~}

\def\ld{\left}
\def\rd{\right}

\def\fr{\frac}
\def\oo{\frac{1}}

\def\del{\delta}

\def\Lam{\Lambda}

\def\sig{\sigma}

\def\LCDM{$\Lambda$CDM}
\def\Planck{\textit{Planck}}

\def\LLTB{$\Lambda$LTB}

\begin{document}

\title{Comment on ``A Supervoid Imprinting the Cold Spot in the Cosmic 
Microwave Background''}

\author{James P. Zibin} \email{zibin@phas.ubc.ca}
\affiliation{Department of Physics \& Astronomy\\
University of British Columbia, Vancouver, BC, V6T 1Z1  Canada}

\date{\today}

\begin{abstract}

   Recently Finelli et al.\ [\href{http://arxiv.org/abs/1405.1555}
{arXiv:1405.1555}] found evidence for a relatively nearby ($z \simeq 0.16$) 
void in a galaxy catalogue in the direction of the cosmic microwave 
background (CMB) Cold Spot.  Using a perturbative calculation, they also 
claimed that such a void would produce a CMB decrement comparable to that 
of the observed Cold Spot, mainly via the nonlinear Rees-Sciama effect.  Here 
I calculate the effect of such a void using a fully general relativistic 
model and show that, to the contrary, the linear integrated Sachs-Wolfe 
effect dominates and produces a substantially weaker decrement than observed.

\end{abstract}

\maketitle


\section{Introduction}

   Recently Finelli et al.\ \cite{finellietal14} (hereafter FGKPS) examined 
the WISE-2MASS galaxy catalogue and found evidence for a void in the 
direction of the cosmic microwave background (CMB) Cold Spot.  They 
furthermore calculated the anisotropy generated by such a void on the CMB 
and claimed that the void could explain most of the Cold Spot decrement, 
mainly due to the nonlinear Rees-Sciama (RS) effect.  In this brief Comment, 
I calculate the effect on the CMB of the best-fit void found by FGKPS, using 
the fully nonlinear spherically symmetric $\Lam$-Lema\^itre-Tolman-Bondi 
(\LLTB) spacetime \cite{omer65}, sourced by dust and $\Lam$, as developed for 
\cite{zm14}.  I consider a standard \LCDM\ universe with \Planck\ best-fit 
parameters \cite{planck13params}.

   I first point out a matter of nomenclature.  Although FGKPS state ``We 
model the underdensity\dots with a \LLTB\ model'', they actually perform 
a perturbative RS calculation using the linearized density (or metric) 
perturbation.  It is not enough for a structure to be spherically symmetric 
to be referred to as a \LLTB\ model---the evolution must also be 
performed using fully nonlinear general relativity, which FGKPS do not do.


\section{The FGKPS profile}

  The spherically symmetric spacetime is described by the metric
\beq
ds^2 = -dt^2 + \fr{Y'^2}{1 - K}dr^2 + Y^2d\Omega^2,
\eeq
for comoving coordinate $r$ and radial derivative $' = d/dr$.  FGKPS adopt 
the curvature function profile
\beq
K(r) = K_0r^2\exp\ld(-\fr{r^2}{r_0^2}\rd).
\label{Kr}
\eeq
It is possible to show that, in the linear regime, the curvature function 
is related to the comoving curvature metric perturbation, $\psi_q(r)$, via
\beq
K(r) = 2r\psi'_q(r).
\eeq
Therefore the FGKPS profile corresponds to
\beq
\psi_q(r) = -\oo{4}K_0r_0^2\exp\ld(-\fr{r^2}{r_0^2}\rd).
\label{psiq}
\eeq
The comoving perturbation $\psi_q$ is time-independent in a dust or dust $+$ 
$\Lam$ background, but is related to the time-dependent zero-shear (or 
longitudinal) gauge metric perturbation, $\psi_\sig$, via
\beq
\psi_\sig(t,r) = \fr{3}{5}g(t)\psi_q(r),
\label{psisig}
\eeq
where $g(t)$ is the growth suppression factor accounting for the suppression 
due to $\Lam$ domination.  The relativistic Poisson equation gives for the 
comoving gauge matter perturbation
\bea
\fr{\del\rho_q(t,r)}{\rho_m}
   &=& \fr{2}{5}\fr{g(t)}{\Omega_m}\fr{\nabla^2}{a^2H^2}\psi_q(r)\\
   &=& \fr{3}{5}\fr{g(t)}{a^2H^2\Omega_m}K_0\ld[1 - \fr{2}{3}\fr{r^2}{r_0^2}\rd]
       \exp\ld(-\fr{r^2}{r_0^2}\rd).
\eea
This agrees with the form of the density contrast in FGKPS [their Eq.~(2)], 
and allows us to identify their density contrast amplitude with
\beq
\del_0 = -\fr{3}{5}\fr{g(t)}{a^2H^2\Omega_m}K_0.
\label{del0K}
\eeq

   FGKPS calculate the RS anisotropy for this profile, and write its central 
amplitude $A \equiv -\del T(\theta = 0)$ as
\beq
A = 51.0\,\mu{\rm K}\ld(\fr{r_0h}{155.3\,{\rm Mpc}}\rd)^3
                     \ld(\fr{\del_0}{0.2}\rd)^2.
\label{RSampl}
\eeq
(Note that here I quote the updated relation from \cite{szapudietal14b}, 
which appears to agree better with the values plotted in Fig.~3 of FGKPS 
than the relation for $A$ given in FGKPS.)  The cubic dependence of $A$ on 
the radius and quadratic dependence on density contrast are well-known 
features of the RS effect in an Einstein-de Sitter (EdS) background (see, 
e.g., \cite{is06,mn09}).

   Finally, FGKPS fit their density profile to the WISE-2MASS galaxy 
distribution, and the predicted anisotropy to CMB data, finding best-fit 
parameters
\bea
\del_0 &=& 0.25\pm0.10,\label{del0bf}\\
r_0h   &=& (195\pm35)\,{\rm Mpc},\label{r0hbf}\\
z_0    &=& 0.155\pm0.037,
\eea
where $z_0$ is the redshift of the void centre.  \eq(\ref{RSampl}) then 
gives central amplitude $A = 158\,\mu{\rm K}$ for the 
best-fit profile, which appears to be consistent with the anisotropy 
profiles plotted in Fig.~3 of FGKPS (assuming sub-dominant integrated 
Sachs-Wolfe (ISW) effect and minor effects due to angular binning).


\section{\LLTB\ calculation}

\begin{figure}
\includegraphics[width=\columnwidth]{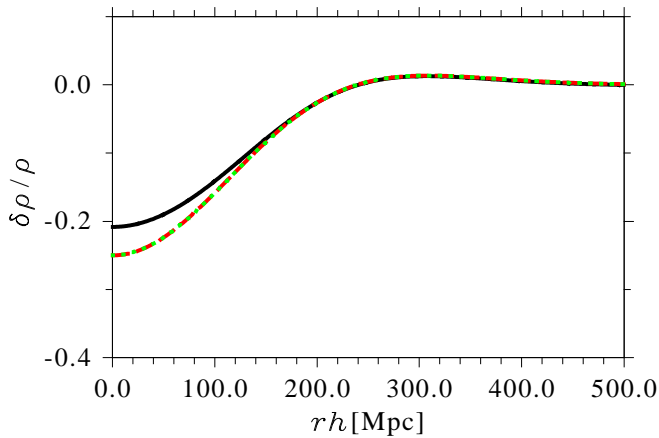}
\caption{Exact \LLTB\ (solid, black curve), linearized \LLTB\ (red, 
dashed), and FGKPS best-fit (green, dotted) three-dimensional density 
contrast.  Here the \LLTB\ model is chosen so the linearized \LLTB\ 
contrast matches that of FGKPS.
\label{linmatch}}
\end{figure}

\begin{figure}
\includegraphics[width=\columnwidth]{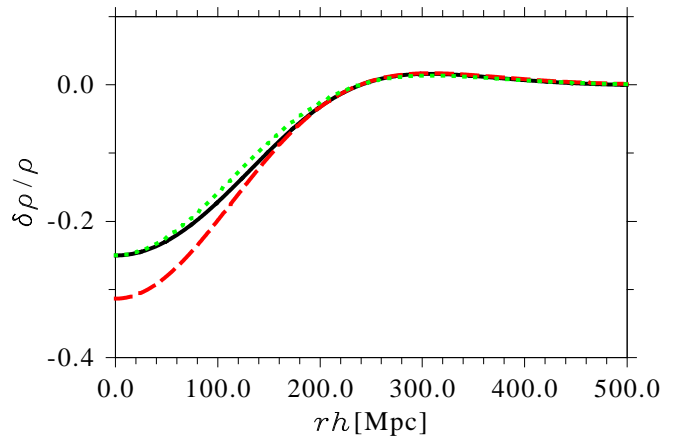}
\caption{As \fig\ref{linmatch}, but here the \LLTB\ model is chosen so the 
exact \LLTB\ contrast matches that of the FGKPS best fit.
\label{nonlinmatch}}
\end{figure}

   In order to calculate with the \LLTB\ spacetime, we must fix the 
curvature function amplitude $K_0$ in \eq(\ref{Kr}) [I will always fix the 
radius $r_0$ to the FGKPS best-fit value \eq(\ref{r0hbf})].  There are two 
ways to do this.  First, we can use \eqs(\ref{del0K}) and (\ref{del0bf}) to 
produce a \LLTB\ profile with the same {\em linearized} density contrast 
as the FGKPS best fit.  The result is plotted in \fig\ref{linmatch}.  In 
this case the nonlinear contrast is smaller than that of the FGKPS best fit, 
as expected since nonlinear growth suppresses underdense contrast.  The 
other approach is to choose $K_0$ so that the {\em exact} \LLTB\ 
contrast matches the FGKPS best fit; this is plotted in 
\fig\ref{nonlinmatch}.  Note that the nonlinear growth changes the 
{\em shape} of the profile; I have chosen the {\em amplitudes} (central 
values) to match.

   Although the difference between these two approaches is small, I 
choose the latter, i.e.\ I choose the exact \LLTB\ contrast to match the 
FGKPS best fit.  This is because this choice corresponds to a larger 
density contrast which will conservatively produce a larger CMB anisotropy 
(and in particular should produce a larger RS/ISW ratio).  This corresponds 
to the value $K_0r_0^2 = -0.00087$.

   Finally I am ready to calculate the CMB anisotropies.  The \LLTB\ 
solution is calculated as in \cite{zm14} by numerically evolving 
Einstein's equations using independent formulations as checks, including 
that descibed in \cite{zibin08}.  I also monitor the constraints and 
compare with LTB (i.e.\ \LLTB\ with $\Lam = 0$) and linear theory in 
the appropriate regimes.  The exact anisotropies are calculated by evolving 
null geodesics from the observer to the last scattering surface, as described 
in \cite{zm11,zibin11}.

   The anisotropies due to the linearized \LLTB\ solution [\eqs(\ref{psiq}) 
or (\ref{psisig})] can be calculated using
\beq
\fr{\delta T}{T} = 2\int\dot\psi_\sig dt
                 + \ld(\fr{5}{3g} - 1\rd)\fr{n_\mu\psi_\sig^{;\mu}}{H},
\label{linanis}
\eeq
where $n_\mu$ is the line-of-sight spatial direction \cite{zs08}.  The first 
term above is the familiar ISW effect, and the second term is the local 
dipole due to the ``bulk flow'' associated with the void.  Even though the 
curvature profile is exponentially damped at large $r$, we will see that 
at the FGKPS best-fit distance $z_0$ the local dipole actually dominates 
over the ISW.

\begin{figure}
\includegraphics[width=\columnwidth]{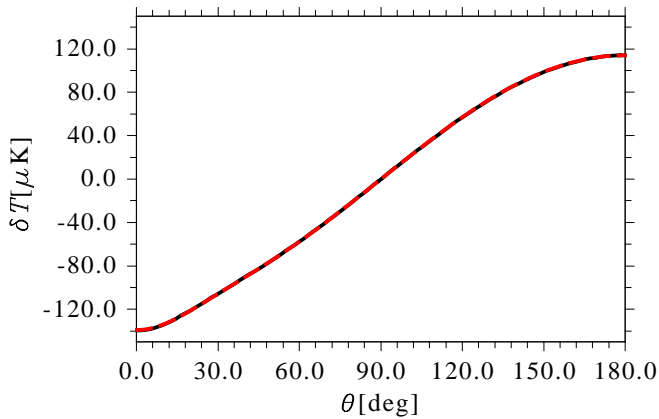}
\caption{Temperature anisotropy calculated exactly using the \LLTB\ 
solution (solid, black curve) and using the linearized relation, 
\eq(\ref{linanis}) (dashed, red).  The two curves are almost 
indistinguishable and are dominated by the local dipole.
\label{delT}}
\end{figure}

   Figure~\ref{delT} shows the anisotropy calculated exactly via raytracing 
in the \LLTB\ spacetime as well as the linearized approximation using 
\eq(\ref{linanis}).  We can see that the local dipolar anisotropy does indeed 
dominate.  But we can also see that the linearized approximation (which 
does not incorporate the RS effect) agrees well with the exact calculation 
(which must include it), which suggests that the ISW dominates over RS.

\begin{figure}
\includegraphics[width=\columnwidth]{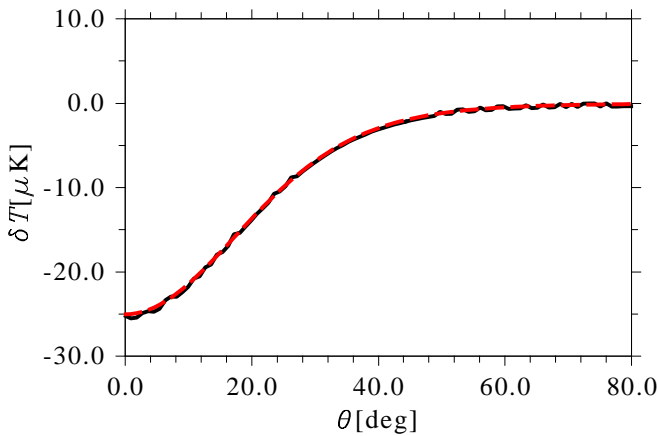}
\caption{As \fig\ref{delT}, but with the (linear) local dipole subtracted 
from both curves.  The ISW clearly strongly dominates the total anisotropy.
\label{delT_nodipole}}
\end{figure}

  We can see this more clearly by subtracting the local dipole contribution 
from both exact and linearized anisotropies, as in \fig\ref{delT_nodipole}.  
Here we see that the anisotropy calculated from the exact \LLTB\ solution 
agrees very well with the linearized, ISW anisotropy.  This demonstrates 
that the RS effect is subdominant for a void of this size, depth, and 
distance.  The central anisotropy is approximately $-25\,\mu{\rm K}$, which 
agrees well with the estimate of $-20\,\mu{\rm K}$ from \cite{szapudietal14}, 
based on the linear ISW effect.

\begin{figure}
\includegraphics[width=\columnwidth]{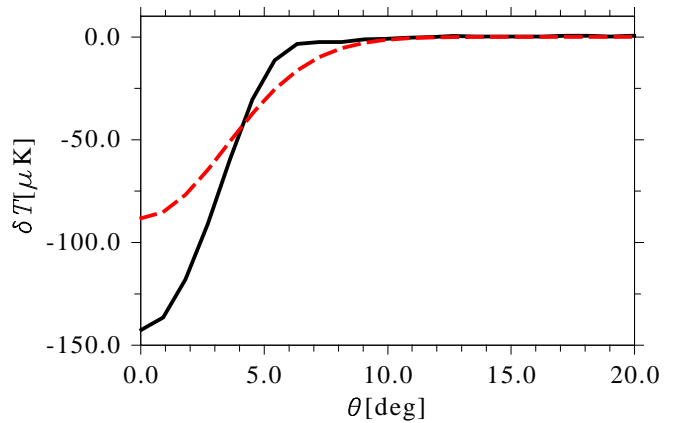}
\caption{As \fig\ref{delT_nodipole}, but for a profile with 10 times the 
amplitude of the FGKPS profile and located at $z_0 = 1$.  The nonlinear 
contribution (RS effect) is now clearly apparent.
\label{delT_K10_z1_nodipole}}
\end{figure}

  Finally, in order to demonstrate that the \LLTB\ code can indeed capture 
the RS effect, I repeat the above calculations for a profile with the same 
width but 10 times the depth of the FGKPS profile, and centred at 
$z_0 = 1$.  The greater depth and distance both ensure a larger RS/ISW 
ratio.  Figure~\ref{delT_K10_z1_nodipole} shows that in this case the 
nonlinear effects are indeed important.

\section{Discussion}

   The question remains as to why the RS prediction of FGKPS, 
\eq(\ref{RSampl}), is so much larger than the nonlinear effect seen in 
the full \LLTB\ calculation.  Although the RS effect is well studied in an 
EdS (dust) background, leading, as already mentioned, to the dependences on 
radius and contrast seen in \eq(\ref{RSampl}), the effect has been much 
less studied in a dust $+$ $\Lam$ background.  However, Ref.~\cite{si08} 
studied the RS effect in this latter case, employing second order, thin 
shell, and fully nonlinear \LLTB\ methods.  They found that the RS 
(nonlinear contribution to the anisotropies) is heavily suppressed in 
the realistic \LCDM\ case with respect to the dust-dominated case, and 
that for comparable void size and depth to the FGKPS profile the linear 
(ISW) anisotropy dominates.

   I have also repeated these calculations using different radial profiles 
$K(r)$, including strongly non-compensated profiles, by fitting the 
profile parameters to closely match the density profile of FGKPS.  In all 
cases I find that the ISW effect dominates over the RS effect and is of 
comparable magnitude to that of the FGKPS profile.  Of course, it is 
entirely possible that the void discussed 
in FGKPS contributes partly to the Cold Spot anisotropy.  But it is clear 
that most of the anisotropy must be sourced elsewhere, since the Cold Spot 
amplitude is considerably larger than $-25\,\mu{\rm K}$: the Cold Spot 
has a deep ``core'' of roughly $5^\circ$ radius and $\sim$$150\,\mu{\rm K}$ 
depth, surrounded by a shallower cold region out to perhaps $10^\circ$ 
radius (see, e.g., FGKPS or \cite{ist10}).  In addition, the 
angular size of the anisotropy shown in \fig\ref{delT_nodipole} (i.e.\ 
$30^\circ$--$40^\circ$) is much larger than that of the observed Cold Spot.  
Although some 
exotic source is always a possibility, a combination of the local void 
with a fluke fluctuation at last scattering might appear most economical.  
Indeed such a scenario was found to be most likely in \cite{inoue12}.



\section*{Acknowledgments} 

This research was supported by the Canadian Space Agency.

\bibliography{comment}

\end{document}